\documentclass{aa}
\usepackage{graphicx}
\usepackage{latexsym}
\usepackage{amsmath}
\usepackage{amssymb}

\newcommand{\MC}{\multicolumn}
\newcommand{\kms}{km\,s$^{-1}$}
\newcounter{qub}
\newcommand{\sunn}{$_{\odot}$}

\begin{document}

\title{Study \mbox{DDO~68}: new evidences for galaxy youth }

\author{%
S.A.~Pustilnik\inst{1} \and
A.L.~Tepliakova\inst{1} \and
A.Y.~Kniazev\inst{2,1}
}

\offprints{S.~Pustilnik  \email{sap@sao.ru}}

\institute{
Special Astrophysical Observatory RAS, Nizhnij Arkhyz,
Karachai-Circassia,  369167 Russia
\and  South African Astronomical Observatory, Cape Town, South Africa
}

\date{Received 24 December 2007; Accepted \hskip 2cm  2008}

\abstract{
DDO~68 is the second most metal-poor star-forming galaxy
(12+$\log$(O/H)=7.14). Its peculiar optical morphology and the data on its
HI distribution and kinematics indicate the merger origin. We use the
photometry of the SDSS $u,g,r,i$ images of DDO~68 to estimate its stellar
population ages.
The available H$\alpha$-images of DDO~68 were used to select several
representative regions without nebular emission.
The analysis of obtained colours was performed via comparison with
the PEGASE2 evolutionary tracks for various star formation (SF) laws,
including the two extremes: instantaneous SF and continuous SF with constant
SF rate.
The $(u-g)$, $(g-r)$ colours derived for all selected regions,
are consistent with a few `instantaneous' SF episodes with ages from
$\sim$0.05 to $\sim$1~Gyr.
Combining the fluxes and colours of visible stellar subsystems with PEGASE2
models, we have estimated the total mass of visible stars in DDO~68
of $\sim$2.4$\times$10$^7$M\sunn.
This comprises only $\sim$6\% of the total galaxy baryonic mass.
All available data do not contradict to the option that DDO~68 is a kind of
very rare candidate
`young' galaxy, whose dominant stellar build-up took place in course of the
recent (with the first encounter $\sim$1 Gyr ago) merger of two very gas-rich
disks.
DDO~68 best approximates on its properties cosmologically young
{\it low-mass} galaxies.
}

\authorrunning{S.A.~Pustilnik et al.}

\titlerunning{Study DDO~68: new evidences for galaxy youth}

\maketitle

\section{Introduction}
\label{intro}

The very metal-poor galaxies were discovered more than 35 years ago. The
first and the most extreme of them, I~Zw~18 (\cite{SS72}), as
well as several similar objects, found in the recent decade and a half (e.g.,
SBS 0335--052~E, \cite{Izotov90}; SBS 0335--052~W, \cite{0335W,VLA},
Lipovetsky et al. \cite{Lipovetsky99}; HS~0822+3542, \cite{Kniazev00},
\cite{SAO0822}; DDO~68, \cite{DDO68}; HS~2134+0400, \cite{HS2134};
UGC~772 and SDSS J2104--0035, \cite{Izotov06}), attract much
attention despite they are very rare in the local Universe. The reason is
that they are thought to
be the best analogs of high-redshift `primeval' gas-rich galaxies.
Some of the most metal-poor gas-rich galaxies are the good candidates for
genuine young galaxies (with ages of stars T$_{\rm stars}$
$<<$13.5 Gyr). Thus, they could represent excellent nearby laboratories to
study in detail the galaxy evolution in the Universe when it was less than
one Gyr old since the Big Bang.

The observational properties of this group, often called eXtremely
Metal-Deficient galaxies (XMD, 12+$\log$(O/H)$\leq$7.65, or Z$<$Z\sunn/10)
or eXtremely Metal-Poor, show
the large diversity, implying that their evolutionary path-ways are different
(see, e.g., Pustilnik \& Martin \cite{NRT}). For many of (still `small'
number)
the well studied XMD blue compact galaxies (BCGs), the colours of outer low
surface brightness (SB) disks are rather red and are consistent with large
(that is comparable to cosmological) ages. However,
for some of the most metal-poor galaxies of this group, the colours of outer
parts are rather blue. The latter several XMD BCGs show no evidences for
`old' stars, and thus are considered as the candidates for local `young'
galaxies. Besides such blue colours, these galaxies have very large gas
mass-fraction $\mu_{\rm g}$=M$_{\rm gas}$/(M$_{\rm gas}$+M$_{\rm stars}$)
(e.g., SBS~0335--052 E and W, with $\mu_{\rm g}$=0.95--0.99, Pustilnik et al.
\cite{VLA,BTA}). The latter also indicates their young evolutionary status.
Recently the prototype XMD galaxy I~Zw~18, based on the Hubble Space
Telescope (HST) data, is shown to possess a sizable population of RGB (red
giant branch) stars (e.g., Tosi et al. \cite{Tosi07}, Aloisi et al.
\cite{aloisi2007}) with ages of T$\gtrsim$2~Gyr. Hence, it is not that
young ($\lesssim$0.5 Gyr) as claimed by Izotov \& Thuan (\cite{IZw18CMD}),
but still can be much younger than the great majority of known galaxies.
In any case, this probable `closure' of I~Zw~18
as a local young galaxy candidate does not close the opportunity of other
very gas-rich XMD BCGs to occur genuine young galaxies. Despite the task to
prove the galaxy youth is a difficult one, the discovery of new candidates,
for which their observational properties do not contradict to the youth
hypothesis and their detailed study is worth of attention, is certainly
exciting. The galaxy DDO~68 can appear one of them.

DDO~68 (UGC~5340) is a galaxy with the second lowest metallicity after
SBS~0335--052~W. Its O/H corresponds to 12+$\log$(O/H)=7.14 (\cite{DDO68};
Izotov \& Thuan \cite{DDO68_IT}) in the new system of relevant atomic
constants suggested by Izotov et al. (\cite{Izotov05}). After correcting
by --0.05~dex, the original value of 12+$\log$(O/H)=7.21$\pm$0.03 from
\cite{DDO68} (in further, PKP) to this new system, the values of O/H from PKP
and Izotov \& Thuan (\cite{DDO68_IT}) are in excellent agreement.
DDO~68 has an unusual optical morphology, with the prominent tidal
tail South of the main body ($\sim$4 kpc across, see, e.g., the deep
$V$-band image in PKP). The absence of visible perturbing neighbours suggests
that this galaxy could be the result of a recent merger.

The HI mapping
of DDO~68 with WSRT (Stil \& Israel \cite{stil02}, Stil \cite{stilthesis})
already hinted on some
peculiarity of its HI morphology (see discussion in PKP). The new high
sensitivity HI mapping of DDO~68 with the GMRT radio telescope (Ekta et al.
\cite{Ekta08}) gave clear detection of two similar `tidal' tails on
the opposite sides of the `body'. The latter evidence for a major
merger of gas-rich components in this object.
From the analysis of the 6-m telescope DDO~68 $V$ and $R$ images, PKP have
concluded that the underlying light is rather blue.
The age estimates of its oldest visible stars resulted in less than 120 Myr
for instantaneous starburst, and of less than 900 Myr - for continuous SF
with constant SFR, both for the case of the standard Salpeter IMF.

In this study we use the independent, high-quality photometry of this object
derived from the Sloan Digital Sky Survey (SDSS) plates in $u,g,r,i$ filters,
to address the question of ages of stellar populations in DDO~68.
In Sec. \ref{reduction}
we describe the used SDSS data and their reduction. In Sec. \ref{results}
the derived magnitudes and colours are presented. Sec. \ref{discussion}
is devoted to the discussion of the obtained results and conclusions. The
accepted distance to DDO~68 of 6.5 Mpc (see below) corresponds to the scale
of 32 pc in 1$\arcsec$.

\section{Observational data and reduction}
\label{reduction}

\subsection{SDSS data description}

The SDSS (\cite{York2000}) is well suited for
photometric studies of various galaxy samples due to its homogeneity, area
coverage, and depth (SDSS Project Book\footnote{
http://www.astro.princeton.edu/PBOOK/science/\\galaxies/galaxies.htm}).
SDSS is an imaging and spectroscopic survey that covers about
one-quarter of the Celestial Sphere. The imaging data are collected in drift
scan mode in five bandpasses ($u, \ g, \ r, \ i$, and $z$; \cite{SDSS_phot})
using mosaic CCD camera (\cite{Gunn98}). An automated image-processing system
detects astronomical sources and measures their photometric and astrometric
properties (\cite{Lupton01,SDSS_phot1,Pier03}) and identifies candidates for
multi-fibre spectroscopy. At the same time, the pipeline reduced SDSS data can
be used for making own photometry (e.g., \cite{Kniazev04}) any project needs.
For our current study the images in the respective filters were retrieved from
the SDSS Data Release 5 (DR5; \cite{DR5}).

Since the SDSS provides users with the fully reduced images, the only
additional step we needed to perform (apart the photometry in round
diaphragms) was the background subtraction. For this all bright stars were
removed from the images. After that the studied object was masked and the
background level within this mask was approximated with the package $aip$
from {\it MIDAS}.  In more detail the method and related programs are
described in \cite{Kniazev04}.
To transform instrumental fluxes in diaphragms to stellar magnitudes, we
used the photometric system coefficients defined in SDSS for the used field.
The accuracy of zero-point determination was  $\sim$0.01 mag in all
filters.

\begin{figure}[hbtp]
   \centering
\includegraphics[angle=0,width=8.5cm,clip=]{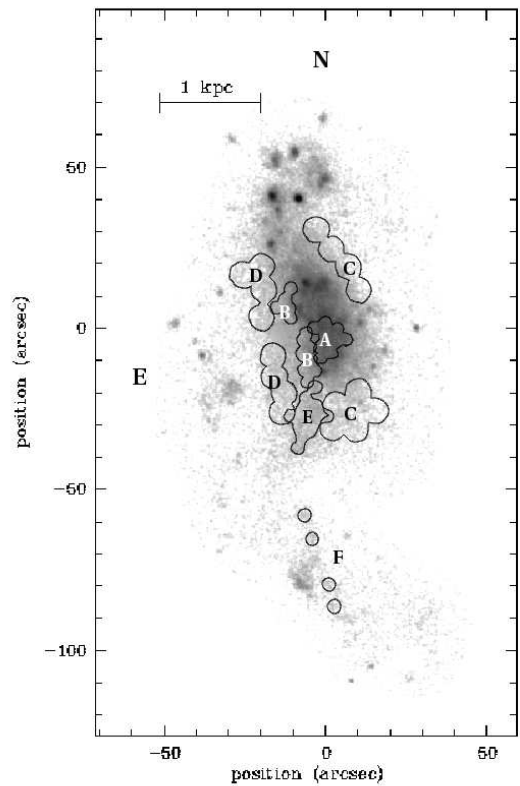}
\caption{
 The SDSS $g$-band image of DDO~68 with contours indicating the
  representative regions without nebular emission, for which the photometry
  data were obtained and analysed. The labels correspond to those in Table
  \ref{t:Photo}. % North is up, East is to to left.
       }
	 \label{fig:FC}
\end{figure}

\subsection{Selection of regions, photometry and control of the derived
magnitudes}

The nebular emission around the sites of current SF in galaxies can
substantially affect the colours of underlying stellar populations (e.g.,
Papaderos et al. \cite{Papa02}; Pustilnik et al. \cite{BTA}). To address the
issue of
stellar ages, one therefore needs in careful selection of regions, where the
contribution of nebular emission in their light is small/negligible. In case
of DDO~68 we based on the net H$\alpha$ images of the galaxy from PKP.
For our analysis we selected only regions
where no H$\alpha$ is detected.

Another important factor to select suitable regions for photometric study
is the understanding of the complex nature of DDO~68 and `a priori' (from
HI imaging analysis and the presence of optical tidal tail) knowledge of its
merger origin. The characteristic timescales of merger are estimated on
the level of 0.5--1 Gyr.
As models of SF in mergers predict (e.g., \cite{springel_etal05}), one
expects two distinct SF episodes: the first during the first encounter, and
the second -- during the coalescence of the merging components.
So, one expects to witness this `young' stellar populations, whose spatial
distribution can be rather inhomogeneous. Whether the old (T$\sim$10 Gyr)
stellar population coexists with this `young' populations in DDO~68? If it
exists, how well this is spatially mixed with the merger starburst products?
Since the galaxy stellar body is rather disturbed and asymmetric, it is not
yet well settled after merging. Therefore one should expect significant
variations of the stellar mixtures in various parts of the galaxy. This poses
additional problems for photometric studies of stellar population ages. It
is thought that the local approach, which deals with several (many) different
regions of relatively small sizes should better account for the expected
differences in the properties of stars along the galaxy body.
For a given depth of the SDSS images, this in turn places respective limits
on the photometry accuracy. Having in mind these circumstances, we
selected the following representative regions, labelled in Fig.~\ref{fig:FC}.

Region A is the nearest to the centre of the main body of DDO~68. Region
B consist of two subregions (B1, B2) on the Eastern side of the main body,
adjacent to the brightest part of DDO~68. Regions C and D correspond to more
distant parts of the galaxy. Each of them consist of two subregions, C1 and
C2, on the western periphery of the main body, and subregions D1 and D2 on
the eastern side. All them are situated on the similar distances from the
galaxy centre. However, the western and eastern parts show systematically
different colours. Therefore we consider them separately.
Region E is situated along the `bridge', connecting the main body and the
southern tail. Region F includes four faint separate subregions (`knots')
in the middle part of this tail. The latter, due to the very low SB and small
total fluxes,
especially in $u$-filter, has the colour uncertainties significantly
larger than for all other regions. We also estimated the mean colours of
several regions on the `distant' western periphery of the main body of
DDO~68 of even lower SB. Due to their very large errors, we did not include
this into following analysis, but will comment them briefly in
Sect.~\ref{discussion}.

Each selected `parent', region shown in Fig.~\ref{fig:FC}, was subdivided into
a number of round adjacent subregions with the radius of 5 or 10 pixels
($\sim$1.2 or 2.4\arcsec). The results of the aperture photometry in round
diaphragms in $u,g,r,i$ filters of all such subregions
were averaged and these average parameters were assigned to the respective
`parent' regions mentioned above. The fluxes of all subregions for each
`parent' region were summed up and transformed to the total magnitudes of
the respective region.

The error budget is as follows. All photometric system related errors are
small since we used the transformation formulae already determined with
the accuracy of $\sim$0.01-0.02 mag in the SDSS database (Navigate Tool).
Due to rather faint fluxes of the majority of measured subregions, the main
error was due to their Poisson noise, estimated directly from the total counts
within each diaphragm. It contributes from a few percent to 20-40 percent,
depending on region and filter. Another factor, affecting in principle the
photometry of rather low SB regions of interest, is the quality of the
background subtraction. The background determination was performed
with the procedure described in detail by Kniazev et al. (2004), where it
was also applied to the SB photometry on SDSS images. This provides very
good quality background subtraction, allowing to perform the photometry on
the SDSS images at the SB levels of 26-29 $g$-mag sq.arcsec$^{-2}$ (when
dealing with rings of sufficiently large radius).

To further perform the control of possible offset in the prepared/subtracted
background, we conducted the standard photometry in circular
apertures with radius of 20 pixels (or 4.8\arcsec) of a dozen stars
% ($g$=16-20)
around DDO~68, which have the SDSS PSF-based (point spread function)
photometry. The obtained shifts between the aperture photometry and the
SDSS PSF photometry appeared to be $\lesssim$0.01 mag in all filters.
This data evidence that our colours for studied regions of DDO~68
are not biased due to background determination.

\section{Results}
\label{results}

\begin{table*}[hbtp]
\centering{
\caption{Average parameters of DDO~86 regions under analysis}
\label{t:Photo}
\begin{tabular}{llllllllc} \hline  \hline
Region  & Distance    & $\mu_g$&$g$-mag & $(u-g)_0$&$(g-r)_0$& $(r-i)_0$& $(V-R)_0$ & Age$^{*}$\\
name    & in $''$/kpc &        &        &$\pm$err&$\pm$err&$\pm$err&                & (Gyr)    \\  \hline \hline
A       &  5/0.16     & 22.30  & 17.18  & 0.69   & 0.03   & 0.02   &  0.13          & 0.12      \\
	&             &        &        & 0.03   & 0.02   & 0.03   &                &            \\
B       &  12/0.38    & 23.14  & 17.91  & 0.88   & 0.08   & 0.08   &  0.14          & 0.3-0.6    \\
	&             &        &        & 0.06   & 0.03   & 0.06   &                &            \\
C       &  26/0.83    & 24.30  & 17.69  & 0.88   & 0.07   & 0.05   &  0.15          & 0.3-0.6   \\
	&             &        &        & 0.08   & 0.05   & 0.08   &                &            \\
D       &  24/0.77    & 24.25  & 17.76  & 1.00   & 0.19   & 0.10   &  0.23          & 0.9-1.1   \\
	&             &        &        & 0.09   & 0.05   & 0.07   &                &            \\
E       &  29/0.93    & 23.81  & 18.68  & 0.67   &-0.02   & 0.10   &  0.11          & 0.09-0.16  \\
	&             &        &        & 0.09   & 0.07   & 0.11   &                &            \\
F       &  72/2.30    & 24.48  & 20.24  & 0.40   &-0.09   & 0.03   &  0.07          & 0.03-0.09  \\
	&             &        &        & 0.20   & 0.20   & 0.35   &                &            \\
\hline   \hline
\MC{9}{l}{$^{*}$ Ages of the regions in assumption of instantaneous SF episode
   with the standard Salpeter IMF.} \\
\end{tabular}
 }
\end{table*}

In Table \ref{t:Photo} we present the results of photometry for the six
selected regions, labelled as A to F in Fig.~\ref{fig:FC}.
In columns 1 and 2 the name of the region and its mean distance
from the DDO~68 centre are given. In columns 3 and 4 we give $\mu_{\rm g}$
- the average SB in $g$-filter and the integrated magnitude of this region
in the same filter. The columns 5 to 7 present corrected for
the Galaxy extinction colours $(u-g)$, $(g-r)$ and $(r-i)$ of the respective
regions with their uncertainties (in the second row). Column 8 presents
the average colour $V-R$ for each region, recalculated from $u,g,r,i$
magnitudes according to the transformation formulae from \cite{Lupton05}.
%as presented at
%http://www.sdss.org/dr5/algorithms/sdssUBVRITransform.html\#Lupton2005.
In the last column the age estimates of these regions are given, which
correspond to the nearest positions (within the ranges of $\pm$1$\sigma$
of their colours) to the PEGASE2 instantaneous SF evolution tracks for
$z$=0.0004 (the nearest to the metallicity of DDO~68) and the standard
Salpeter IMF. This numbers will be discussed in more detail in
Sect.~\ref{discussion}.

The range of SB, probed by our selected regions is about 2.2 mag. The
brightest one, near the centre, corresponds to
$\mu_{\rm B}\sim$22.5~mag~sq.arcsec$^{-2}$, while the regions near the
middle of the southern
tail correspond to $\mu_{\rm B}\sim$24.7~mag~sq.arcsec$^{-2}$.
The colours of the majority regions are rather blue in $(g-r)$
($\lesssim$0.10),
and moderate in $(u-g)$ (0.7-0.9) and $(r-i)$ (0.05-0.12).
The large errors for colours of Region F are related to its very low signal.
In particular, in $i$-filter the object flux is 1.5 times lower than in $r$,
while the Poisson noise due to the higher level of background is 25\% higher
than in $r$.
The lowest SB regions on the western periphery of DDO~68, for which we got
the estimates of colours, have the mean
$\mu_{\rm B}\sim$26.3~mag~sq.arcsec$^{-2}$.

\section{Discussion and conclusions}
\label{discussion}

\subsection{Comparison with the model evolutionary tracks}

\begin{figure*}[hbtp]
   \centering
 \includegraphics[angle=-90,width=15cm,clip=]{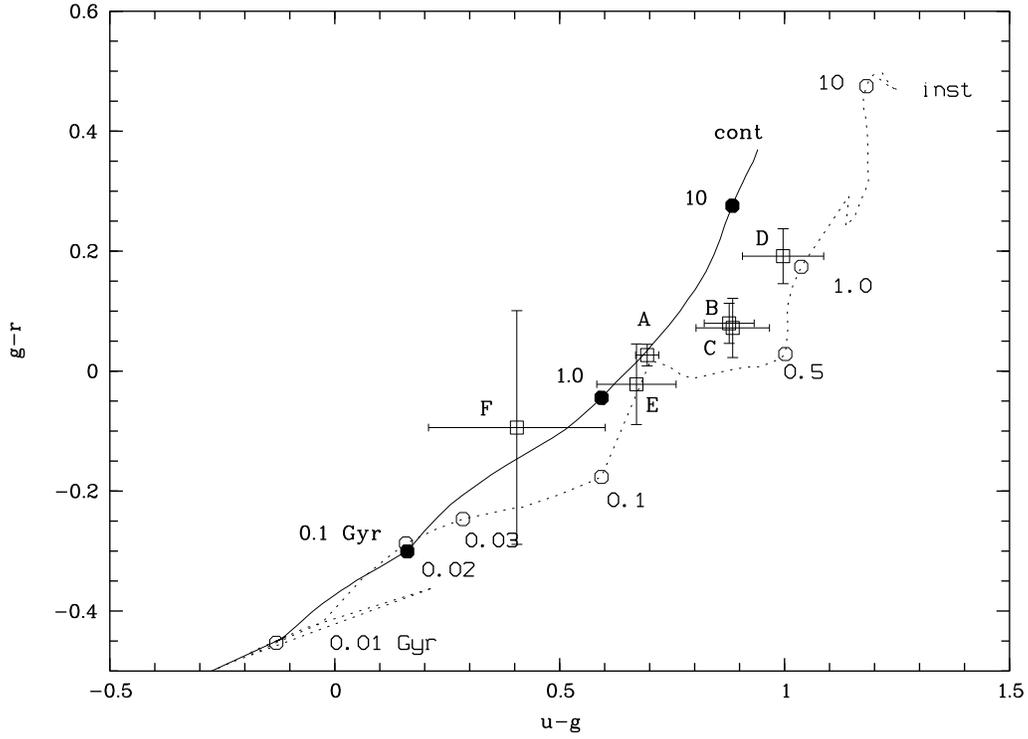}
   \caption{
Two-colour $(g-r)_0$ vs $(u-g)_0$ diagram with the theoretical tracks from
   PEGASE2
   for evolving stellar populations with the standard Salpeter IMF, for
   instantaneous (dashed) and continuous (solid line) SF laws. Filled and
   empty hexagons along the tracks, with the respective
   numbers,  correspond to ages since the beginning of SF (in Gyr).
   The observed colours for the regions discussed in the text are shown
   by empty squares with their $\pm1\sigma$ error bars and are labelled
   by the respective letters (A-F).
}
	 \label{fig:ugr}
\end{figure*}

\begin{figure*}
%   \vspace*{1cm}
   \centering
 \includegraphics[angle=-90,width=15cm, clip=]{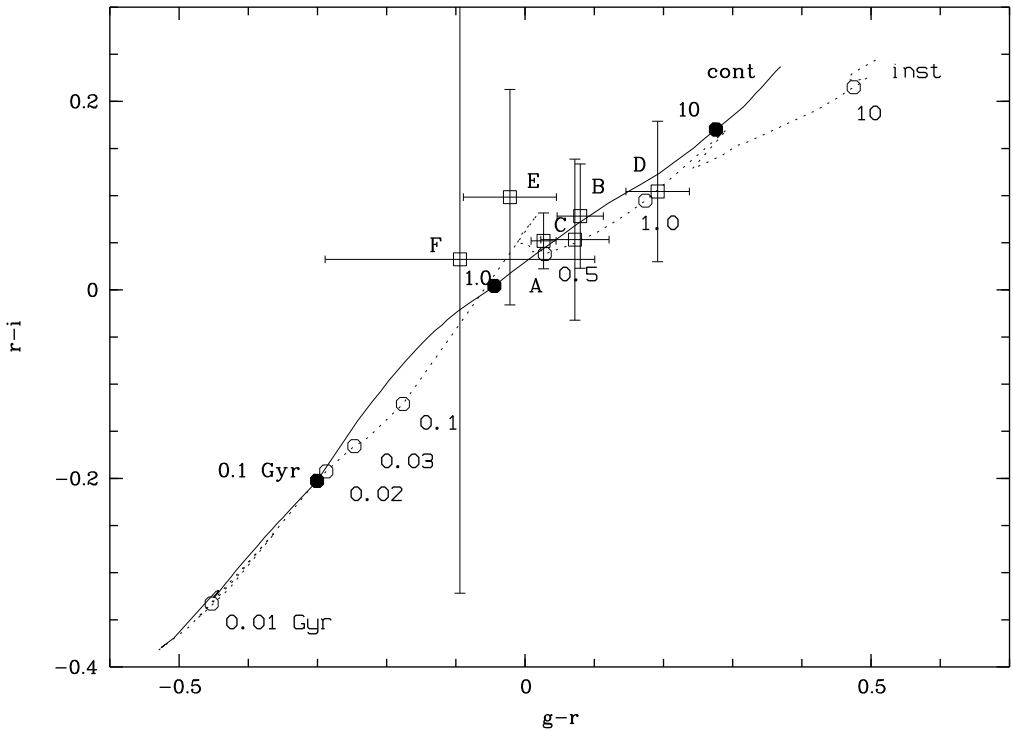}
   \caption{
Same as for the previous figure, but for $(r-i)_0$ vs $(g-r)_0$ diagram.
}
	 \label{fig:gri}
 \end{figure*}

Since the original goal of this work was to estimate ages of stellar
populations in DDO~68, we confront the derived colours in its different parts
with the model tracks from the PEGASE2 package (Fioc \& Rocca-Volmerange
\cite{pegase2}) for metallicity $z$=0.0004.
In fact, the photometric systems
($u^{\prime}$,$g^{\prime}$,$r^{\prime}$,$i^{\prime}$,$z^{\prime}$)
used for calculations of PEGASE2 evolutionary tracks and ($u,g,r,i,z$) used
in the real SDSS observations are slightly different. We applied the
transformation formulae from \cite{Tucker06} in order to correct
theoretical values to ($u,g,r,i,z$) system.
In Fig.~\ref{fig:ugr} and \ref{fig:gri} we plot the model tracks of colour
evolution in $(g-r)$ versus $(u-g)$ and $(r-i)$ versus $(g-r)$ diagrams.
%for the standard Salpeter
%IMF with lower amd upper limits of 0.1 and 120 M\sunn.
The two tracks, shown by the solid and dotted lines, represent the colour
evolution for continuous SF with constant SFR and for instantaneous SF
episode, as two extremes of all possible SF histories. The standard
Salpeter IMF with lower and upper limits of 0.1 and 120 M\sunn\ was accepted.
The hexagons on evolutionary tracks with the respective numbers mark ages
since the beginning of SF (in Gyr).
%We do not show here (due to crowding the figures) the track, corresponding
%to one of intermediate cases, when SF took place during the first 0.1 Gyr and
%then stopped. This case corresponds to an enhanced SF episode, induced by
%the strong interaction/merger.
The positions of observed extinction-corrected colours (with E($B$-$V$)=0.018,
Schlegel et al. \cite{Schlegel98}) in selected regions of DDO~68, summarised
in Table~\ref{t:Photo}, are also shown in the figures. Below we discuss in
more detail the possible interpretation of these colours, region by region.

The observed $(u-g), (g-r)$ for the `central' Region A  are rather blue and
equally well correspond to an instantaneous starburst with the age of
$\sim$0.12 Gyr or to continuous SF episode with the age of $\sim$1.6~Gyr.
The colours of two adjacent subregions belonging to Region B (the periphery of
the main body) are redder; they fall in this diagram between the two tracks.
The nearest (in probabilistic sense, accounting the error bars in $(u-g)$ and
$(g-r)$) to the observed colours the points on the instantaneous SF track
correspond to ages of $\sim$0.5~Gyr. The continuous SF track is somewhat more
distant and its nearest point corresponds to ages of $\sim$2.5~Gyr.
Region C is in average 2 times more distant from the centre, than Region B,
and its $g$-filter SB is $\sim$1.1-mag fainter. Their $(g-r)$ colours differ
only by 0.01 mag, while the $(u-g)$ colours are the same.
Since their colours are very close, the respective age estimates for Region C
are the same as for Region B.
Region D, situated on the eastern periphery, have the reddest $(u-g)$ and
$(g-r)$ colours, which fall sufficiently close to the track for instantaneous
SF, with the most probable age of $\sim$1 Gyr (with the range corresponding
to $\pm$1$\sigma$ of $\sim$0.9 to $\sim$1.1 Gyr). The probability that the
colours of Region D relate to the track with continuous SF is significantly
lower. The nearest points of this track correspond to ages of $\sim$6~Gyr.

Region E is situated in the `bridge' between the main body and
the tail, stretching to the South. Its $(u-g)$ and $(g-r)$ colours are
somewhat bluer than those of Region A, and fall very close to the track of
instantaneous SF.
The formal range of the ages, corresponding to $\pm$1$\sigma$ range of its
colours, is from 0.09 to 0.16 Gyr, with the most probable estimate of
$\sim$0.11 Gyr. The nearest points on the continuous SF track correspond to
the ages of $\sim$1.4~Gyr

We also measured the colours of four resolved `knots' of Region F,
situated in the middle part of the tail, on the both sides of
the `ring-like' HII region (No.~7 in PKP nomenclature). Due to their low SB
and the small sizes, the S-to-N ratio
for this region is lower than for the others. Its face value $(u-g)$, $(g-r)$
colours fall close to the points on the track with continuous SF with
ages of $\sim$0.5 Gyr.
However, due to very large errors, the option of instantaneous SF with ages
of 0.03--0.09 Gyr is an acceptable alternative. Taking into account
the tracers of the current SF in the close environment (the ring-like HII
region), we favour the latter option of a relatively recent SF in the tail
as a more realistic.

Return again to the interpretation of colours for regions B and C, which
deviate from the track for instantaneous SF. It is clear, that the stellar
population with ages of $\sim$1~Gyr, the most clearly seen in Region D,
can certainly contribute to the colours of other regions as well. In this
aspect, the colours of regions B and C the most naturally are explained as
colours of a composite population, that is those which result from the mixture
of radiation of `young' (T$\sim$0.12--0.14~Gyr) and `old' (T$\sim$1~Gyr)
stars. The track in $(u-g)$, $(g-r)$ diagram, corresponding to varying
ratio of component fluxes in such a mixture, will join the points with
respective ages, belonging to the instantaneous SF track. In order not to
crowd Fig.~\ref{fig:ugr} we do not show this composite track.
The observed colours of regions B and C best correspond to the mixture
with $g$-filter flux ratio of 1.9:1, corresponding to the mass ratio of
M(0.14~Gyr)/M(1.0~Gyr)=0.5.

The errors in $i$-filter for all regions appear significantly larger, than
in others (due to the combined effect of smaller flux and higher sky noise).
Therefore, $(r-i)$ colour appears unuseful in the further constraints of the
old population age estimates. We just notice that the $(r-i)$ colours for all
regions are consistent within their uncertainties with the age estimates
derived from $(u-g)$ and $(g-r)$  colours.

It is worth to comment the colours of the outermost regions we tried to
measure. They are as follows: $(u-g)$=0.87, $(g-r)$=0.12 (not in
Fig.~\ref{fig:ugr}). These are very low SB regions to the west of the main
body of DDO~68. Their average distance from the centre is of $\sim$35\arcsec\
and the mean SB $\mu_{\rm B}\sim$26.3~mag~sq.arcsec$^{-2}$. The errors for
the colours of $(u-g)$ and $(g-r)$ (0.26 and 0.15 mag, respectively) are too
large to uniquely distinguish between tracks for instantaneous or continuous
SF. However, the measured colours themselves are close to those in regions B,
C and D. This implies that within the cited uncertainties, the colours of the
outermost parts of DDO~68 do not contradict the estimates of their ages
of $\lesssim$1~Gyr.

\begin{figure*}
 \includegraphics[angle=-90,width=15cm, clip=]{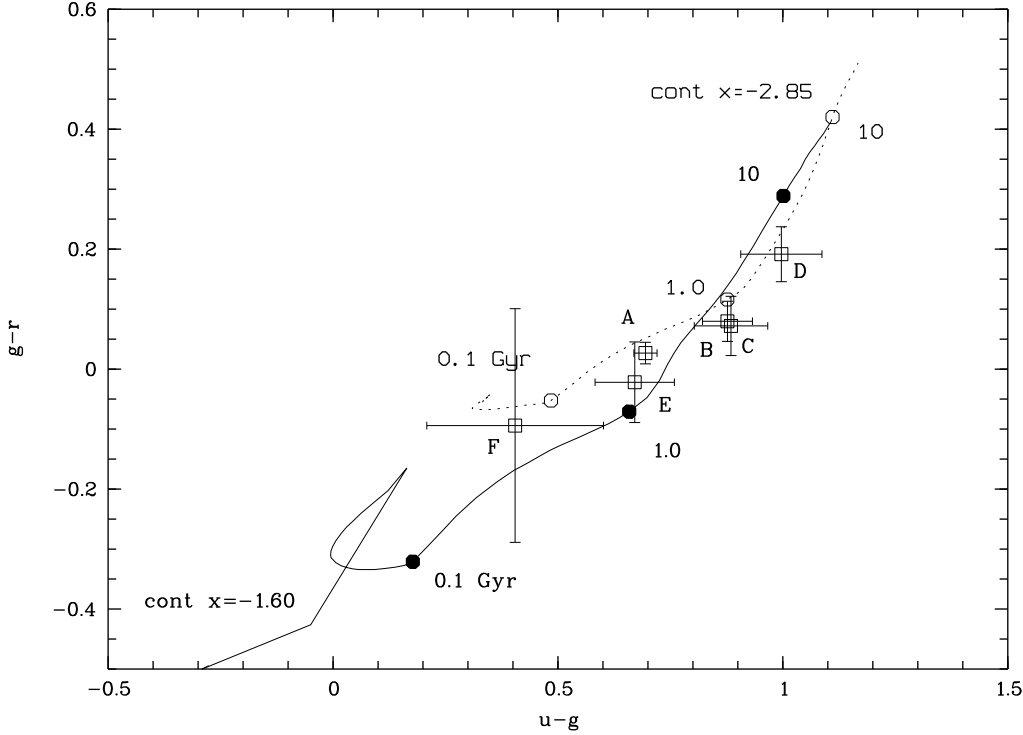}
   \caption{
Two-colour $(g-r)_0$ vs $(u-g)_0$ diagram with theoretical tracks from
PEGASE2  for evolving stellar populations with two variants of the
bottom-heavy IMF with slopes of $x$=--2.85 and --1.60. Only tracks
for continuous SF law with constant SFR are shown. All symbols for model
and observed colours are the same as in Fig.~\ref{fig:ugr}.
}
	 \label{fig:bottom}
 \end{figure*}

In addition to the mentioned above extreme SF laws, we compared the observed
colours with the track for the intermediate case, which better corresponds to
SF in course of merger. Namely, we considered the colour evolution for SF with
constant SFR during the first 0.1 Gyr, which then ceased. This track on
$(u-g), (g-r)$ diagram (not shown due to crowding) follows naturally for the
first 0.1 Gyr that for continuous SF. For ages of $\gtrsim$0.2 Gyr (where
most of our observed colours fall) it follows very close to that of
instantaneous SF. For intermediate ages this track goes approximately in the
middle between the two extremes.
Therefore, summarising this point, we conclude, that for the standard
Salpeter IMF, the observed colours of
DDO~68 representative regions well correspond to `short' (with the duration of
$\lesssim$0.1 Gyr) SF episodes with ages of $\sim$0.12--0.14 and
$\sim$1.0 Gyr.

Below we discuss the alternative options for interpretation of the observed
DDO~68 colours with the use of non-standard IMFs. As noted below, one of the
options for DDO~68 progenitors can be LSB galaxies. The integrated colours of
LSBGs are found to be in the broad range from blue to red. It is important
for further that in the substantial fraction of LSBGs galaxies faint unusually
red halos are discovered (e.g., \cite{Red_halos}). One of the possible
interpretations of this phenomenon is the result of so called bottom-heavy
IMF, in which the fraction of massive stars is much lower than in the
standard Salpeter IMF with the slope of $x$=--1.35. To explain the
unusually red colours of these halos the IMFs with slopes of --2.85 to
--3.50 were suggested (Lee et al. \cite{Lee04}, \cite{Red_halos}).
However, incorporation of such steep IMFs into models of chemical evolution
and confronting the model predictions with observational data on O/H and N/H
for large LSBG sample (\cite{mattsson2007}) indicates that very steep IMFs
($x$=--2.85) do not agree with observations. Reasonably good agreement is
obtained for IMF with $x$=--1.60.

In Fig.~\ref{fig:bottom} we compare $(u-g)$, $(g-r)$ PEGASE2 tracks for
continuous SF with constant SFR and $z$=0.0004 for cases of IMF with the
slopes of $x$=--2.85  and $x$=--1.60, with the observed colours of DDO~68
regions.
The track for $x$=--2.85 goes sufficiently close to the colours of all
regions.
The reddest colours (Region D) correspond to ages of less than 2~Gyr, while
those for the other regions -- to ages of less than 1~Gyr. In the diagram
$(g-r), (r-i)$ this track goes systematically above the observed colours.
The track for continuous SF with $x$=--1.60 goes significantly closer to
colours of regions B, C and D, than the same track with the standard IMF
($x$=--1.35). The colours of these regions equally well correspond to
the track with instantaneous SF and the standard IMF and to continuous SF
track for IMF with $x$=--1.60. For the latter case, the nearest points
correspond to ages of $\sim$2~Gyr for regions B and C, and to ages of
$\sim$5~Gyr for region D. In the diagram $(g-r), (r-i)$ the latter track goes
very close to the Salpeter IMF tracks, so no one of them is preferable.

Summarising, we conclude that formally the track with continuous SF and
IMF with $x$=--1.60 gives more or less acceptable agreement with the
observed colours of DDO~68 regions. However, combining the derived age
estimates with the picture emerged from the optical and HI morphology and gas
kinematics, it is difficult to suggest a realistic model, in which in
different regions of the galaxy the continuous SF took place with the
significantly different timescales (from $\sim$1 till $\sim$5 Gyr) and in the
same time there were no appearance of SF, related to the recent tidal
disturbances.

Therefore it is more correct to pose the question as follows. If a fraction
of light observed in the studied DDO~68 regions is caused by the contribution
of stars, whose evolutionary status corresponds to the continuous SF track
for IMF with $x$=--1.60, then what mass estimates of this population will
not contradict to all other observational data. Below we give the estimate
of stellar mass for the case of SF with the standard Salpeter IMF, as well as
for the option when the reddest stars are from the population with
the non-standard IMF.

\subsection{The stellar mass and gas mass-fraction in DDO~68}

To derive an estimate of the stellar mass of DDO~68, the next parameters
are used, as determined on the photometry of its SDSS image.
First, this is the total $g$-filter magnitude after removal of the most
evident foreground stars: $g_{\rm tot}$=14.42. Second, the $g$-filter
magnitude of the central `plateau' (including region A), in which the main
contribution comes from the `young' population and HII regions, with the
addition of the light from the other SF regions (including the N and S
ring-like HII regions and others) $g_{\rm plat}$=15.56. Third, the $g$-filter
magnitude of the rest (outer) parts, with the colours typical of regions B
and C, which is derived through the difference of the total flux and that
of the `plateau', $g_{\rm outer}$=15.23. Then, the estimate of stellar mass
is performed, suggesting that in the outer parts of DDO~68  the populations
with ages of 0.14 and 1 Gyr contribute in the same proportion as in that
derived for regions B and C, while for `young' population region the main
contribution to the light comes from stars with the ages of $\sim$0.14~Myr.

Then, from the mentioned above the PEGASE2 evolutionary tracks, with
corrections to the SDSS filter system according to \cite{Tucker06},
we obtain luminosities per 1~M\sunn\ for respective ages. Then, calculating
from visible magnitudes the respective absolute ones (with the distance
module $\mu$=29.06 mag)\footnote{We accepted for the estimates of masses of
stars and gas the distance $D$(DDO~68)=6.5~Mpc, as in paper by PKP. However,
it can be significantly larger since the collection of
all independent distance measurements of galaxies in this volume, presented
by Tully et al. (\cite{Tully07}), indicates large negative peculiar
velocities.}
we derive masses for both stellar populations
M(0.14~Gyr)=1.1$\times$10$^7$M\sunn\ and M(1~Gyr)=1.3$\times$10$^7$M\sunn,
with the total stellar mass of M$_{*}$=2.4$\times$10$^7$M\sunn.
The minimal gas mass (without molecular and ionised gas components)
derived from HI-flux of Ekta et al. (\cite{Ekta08}), with account for the
mass fraction
of He, is M$_{\rm gas}$=1.33$\times$M(HI)=38$\times$10$^7$M\sunn.
From this, the gas mass-fraction is
$\mu_{\rm gas}$=M$_{\rm gas}$/(M$_{\rm gas}$+M$_{*}$)$\sim$0.94.

For the alternative IMF, with the slope of $x$=--1.60, the colours of region
D are interpreted as a result of continuous SF during the last 5 Gyr.
Suggesting that the contribution of this kind of stellar population in the
light of DDO~68 is the same, as for instantaneous SF population with ages of
$\sim$1~Gyr from the previous case, the estimated mass of this component is
as follows: M(5~Gyr)=2.0$\times$10$^7$M\sunn. If all the light from the bluer
components is assigned to the population formed in instantaneous SF episode
with ages of $\sim$0.14~Gyr (as in the previous case), then its mass estimate
will be again of M(0.14~Gyr)=1.1$\times$10$^7$M\sunn. So, the total stellar
mass will be of M$_{*}$=3.1$\times$10$^7$M\sunn. The respective to this case
value of $\mu_{\rm gas}$ will be $\sim$0.92.

\subsection{Possible model of DDO 68}

The estimates of the ages and the mass of stellar population in DDO~68,
derived in the previous sections, along with the recent GMRT results on
its HI density and velocity field (Ekta et al. \cite{Ekta08}), shed new light
on the nature of this unusual galaxy. Below an attempt is made to draw
an empirical model of DDO~68 which accounts for all available to-date
observational findings.
There are the following main facts which should be explained in the frame
of one scheme.

First, all available spectral data on DDO~68 are consistent with the
conclusion that its ISM metallicity is everywhere at the level of Z\sunn/30
(PKP, Izotov \& Thuan \cite{DDO68_IT}).
Second, its morphology, both in the optical range and in HI 21-cm line is
strongly disturbed. The nearest potential `disturber' is the dIrr galaxy
UGC~5427, situated at the projected distance of $\sim$200 kpc. Its absolute
$B$-magnitude is by 0.5 mag fainter than for DDO~68. It is very unlikely that
this galaxy caused such disturbed morphology and kinematics of DDO~68.
Moreover, HI data clearly evidence for the recent major merger of two
gas-rich objects in this galaxy.
Third, from the photometry of several representative regions of DDO~68, free
of nebular emission, there are no indications for the existence
of stellar populations with ages larger than $\sim$0.1-1.0 Gyr, if their
colours are compared with evolutionary tracks for the standard Salpeter IMF.
In case of a `bottom-heavy' IMF, the stars with ages of $\lesssim$5~Gyr can
give the contribution of $\sim$5\% to the total baryon mass of the galaxy.
Forth, the gas mass-fraction in DDO~68 is unusually high for late-type
galaxies, comprising of $\sim$0.92--0.94.

Several immediate conclusions follow from the above consideration.
%\begin{enumerate}
%\item
{\bf 1.} The `large' baryon mass ($>$4.0$\times$10$^8$M\sunn) and several
times larger the total dynamical mass imply that the metal loss in DDO~68 was
insignificant during previous evolution. Therefore, its observed extremely
low ISM metallicity is the result of very slow astration and the respective
very low rate of metal production (if at all) in DDO~68 progenitor(s).
%\item
{\bf 2.} The above conclusion implies two possible options for DDO~68
progenitors.
They were either old, very slowly evolving (Very) Low Surface Brightness
galaxies (as suggested for I~Zw~18, e.g., by Legrand et al.
\cite{Legrand2000}),
experienced recently merger. Or these were a kind of protogalaxies
(`dark galaxies'),
in which no stars have formed until the `recent' SF episode, induced by
merger. There is also a merger option of a VLSB and a protogalaxy,
%\item
{\bf 3.} For any of the options, all `young' stellar populations, with ages
of $\lesssim$0.1-1 Gyr, formed in the SF episodes related to `galaxy'
collisions: starting after the first close encounter and then after several
hundred Myr, in course of the subsequent merger of two very gas-rich objects
with comparable masses.
%\item
{\bf 4.} The main difference between these possible path-ways is the
presence of the sizable amount of old stars (with ages of $\sim$5-13 Gyr),
if the progenitor was a LSB dwarf with very low average SFR in previous
epochs.
%\end{enumerate}

DDO~68 is not a typical BCG, since its SF activity is not that
strong. The EWs(H$\alpha$) in the main body are rather small that implies
a decaying SF episode with ages more than $\sim$10-15 Myr. The strongest
appearance of the recent starbursts (with ages of $\sim$4--7 Myr) are seen
at the galaxy periphery, in two ring-like structures, the Northern `ring'
and the similar Southern feature, falling in the middle of the tail (PKP).
Both of them (judging from their morphology) can be of a secondary origin.
Namely, they can be triggered by shells, generated by previous starbursts
near the centres of these `rings'. The age estimates of stars near the galaxy
centre and in the tail, presented above, indicate that the main SF activity
probably ceased from several tens to a hundred Myr ago.

Combining all available data on DDO~68, first of all, from PKP, Ekta et al.
(\cite{Ekta08}) and this work, its the most likely model can be drawn as
follows.
Two very gas-rich objects ($\mu_{\rm gas}\sim$0.95--1.0), each with the
characteristic baryon mass of $\sim$2$\times$10$^8$~M\sunn, have collided
at first time about $\sim$1~Gyr ago. This first encounter have induced the
significant disturbances in both `galaxies', which resulted in the first
SF episode, with the formed stellar mass of $\lesssim$1$\times$10$^7$~M\sunn\
($\lesssim$3~\% of the total baryon mass). During the second collision of
these two objects, they have merged and caused a larger disturbance of gas,
that generated two apparent HI tidal tails (Ekta et al. \cite{Ekta08}) and
induced the `central' starburst. This starburst
resulted in formation of the second part of DDO~68 stellar mass with ages of
$\lesssim$1~Gyr, with the total stellar mass of stars of
$\sim$1$\times$10$^7$~M\sunn\ for ages of $\lesssim$0.14~Gyr.
The SF in tails was delayed and proceeded due to the gas collapse in clumps
only in the recent epochs. The results of this SF are visible as the `younger'
stellar population in the middle of the southern tail. The latest episodes of
SF in tails with ages of several Myr are seen
as the northern and southern `rings' of HII regions, discussed by PKP.
The colours of the outermost regions, located, in particular in very LSB
western periphery, at the distances of $\sim$1~kpc from the centre, despite
to rather large errors,  agree with the hypothesis that they also formed
during the last 0.5-1 Gyr. Summarising, we suggest that DDO~68 is a merger
with the dominating `young' low-metallicity stellar population, and a likely
very rare candidate to a genuine young galaxy in the local Universe.
However, the question of the real IMF for DDO~68 stars is a crucial one
for determination of its evolutionary status.

For any of the possible path-ways, DDO~68 is one of the best analogs of the
high-redshift low-mass young galaxies. Thus, the modelling and interpretation
of its properties would help in understanding the issue of interactions of
low-mass galaxies in the early Universe. Besides, DDO~68 is one of the nearest
galaxies with that extremely low ISM metallicity. Its young massive stars
with metallicity Z$\sim$Z\sunn/30 are the best targets for the next
generation of giant optical telescopes to study directly their evolution
(see also Kniazev \& Pustilnik \cite{IUAS232}).

\subsection{Prospects and models}

The properties of DDO~68 are quite unusual. At the current level of our
knowledge it remains a real candidate for a `young' local galaxy. There are at
least two options to check the presence of older stellar populations.
The first, ground-based one, infers the deeper surface photometry of outer
parts in order to apply the analysis, similar to the presented above, and
the comparison with model tracks.
The deep photometry in $U$ or $u$-bands is expected to be useful in order to
disentangle the degeneracy of ages and SF laws.

An alternative is a space-based option, with
the deep imaging and subsequent analysis of CMD diagrams for individual stars.
This assumes the use of, e.g., the HST or its successor opportunities. The
latter will require two-band images of selected  regions in DDO~68 with the
limiting magnitudes in $V$-band of 27--28, in order to well registrate the
tip of RGB (M$_{\rm V}$=--4.0), if this population is present.
The latter estimate accounts for the fact that the current value of DDO~68
is rather uncertain and can be as large as $\sim$10 Mpc,
corresponding to the distance module of $\mu\sim$30.

It is worth to notice the importance of detailed studies and modelling of
DDO~68 and several similar objects, since their properties indeed best
approximate the properties of young high-redshift low-mass galaxies.
Some of their observed parameters probably already indicate limitations of
models related to such objects. In particular, one can mention the prediction
of model by Elmegreen et al. (\cite{elmegreen93}) of the
significant gas `agitation', with the increase of its velocity dispersion
in course of merger by several times. Despite the early DDO~68 HI data by
Stil (1999) indicated that the HI velocity dispersion was elevated by a
factor 2 to 3 in several regions (see discussion in PKP), the recent
observations of Ekta et al. (\cite{Ekta08}) put upper limits on this
parameter.
Namely, the velocity dispersion is elevated of no more than by a factor
of 1.3-1.4 of its standard value of 7-8~\kms.

Another interesting point relates to N-body simulations of almost purely
gaseous disk mergers (Springel et al. \cite{springel_etal05}, Springel \&
Hernquist \cite{springel05}).
Their models, with the most updated SF feedback prescripts, predict that
after the coalescence of such disks, the resulting object will retain the
properties of a disk galaxy. However, the expected in models the gas
mass-fraction in the merger product appears only of 20--30~\%. While
DDO~68 is most probably the result of very gas-rich merger, its estimated
gas mass-fraction of $\sim$92--94\% appeares in a drastic
contrast with the above-mentioned model results.
Of course, some of the input model parameters differ from the observed ones.
In particular, the masses of the involved merger components are much larger
than that of DDO~68. This, as well as the SF feedback prescripts, can affect
the parameters of the resulting merger.
Therefore, we emphasise that similar models for low-mass gas-rich objects
are very actual, since there exist real objects with which such models can be
confronted.

\subsection{Conclusions}

Summarising the results of this work and the discussion of all available
data on DDO~68, we draw the following conclusions:

\begin{enumerate}
\item
We obtained from the photometry of DDO~68 SDSS $u,g,r,i$ images the colours
of its several representative regions, which are not contaminated by the
nebular emission of the current SF episode.
The comparison of their $(u-g), (g-r)$  colours with the model PEGASE2 tracks
for the standard Salpeter IMF indicates `young' stellar populations,
with ages in the range from tens Myr to $\sim$1~Gyr. These ages are
consistent with the formation of stars in these regions in course of the
recent merger event.
\item
Counting the main part of visible stellar light and its observed colours, we
estimated (for the standard Salpeter IMF and the metallicity of $z$=0.0004)
the total mass of stars in DDO~68 of $\sim$2.4$\times$10$^{7}$M\sunn.
The latter comprises $\lesssim$6\% of the total baryon mass that includes
the mass of atomic hydrogen and helium and the mass of stars.
\item
A deeper surface photometry of DDO~68 periphery (including $U$-band data)
will be useful to check the presence of older stellar component(s). The
space-based photometry of the resolved stellar populations with the limiting
magnitudes of $V$$>$27--28 mag. could probe the presence/absence of RGB stars
and help to better understand the origin of this unusual system.
The new data are also necessary to put limits on its possible IMF.
\item
Summarising all available observational data, we conclude that DDO~68 can be
one of the youngest galaxies in the local Universe, with the ISM and the
{\it massive} star metallicity near the bottom of known to date metallicity
distribution for gas-rich galaxies. This opens a good opportunity
to study directly the SF and evolution of massive stars at such low
metallicities.
\end{enumerate}

\begin{acknowledgements}
The study of SAO authors was supported by the RFBR grant No. 06-02-16617.
The authors acknowledge the high-quality photometric data and the related
information available in the SDSS database, used for this study.
The Sloan Digital Sky Survey (SDSS) is a joint project of the University of
Chicago, Fermilab, the Institute for Advanced Study, the Japan Participation
Group, the Johns Hopkins University, the Max-Planck-Institute for Astronomy
(MPIA), the Max-Planck-Institute for Astrophysics (MPA), New Mexico State
University, Princeton University, the United States Naval Observatory, and
the University of Washington. Apache Point Observatory, site of the SDSS
telescopes, is operated by the Astrophysical Research Consortium (ARC).
\end{acknowledgements}

\end{document}